\documentclass[11pt,twoside]{article}
\usepackage{asp2010}

\resetcounters

\begin{document}

\title{Overview of the SOFIA Data Cycle System:  An integrated set of tools and services for the SOFIA General Investigator}
\author{R.~Y. Shuping$^{1,2}$,William D. Vacca$^2$, Lan Lin$^2$, Li Sun$^2$, \& Robert Krzaczek$^3$
\affil{$^1$Space Science Inst., 4750 Walnut St., Suite 250, Boulder, CO  80301}
\affil{$^2$USRA-SOFIA, NASA Ames Research Center, N-232, Moffett Field, CA  94035-0001}
\affil{$^3$Carlson Center for Imaging Science, Rochester Inst. of Technology, 54 Lomb Memorial Dr.,
Rochester, NY 15623}
}

\begin{abstract}
The Stratospheric Observatory for Infrared Astronomy (SOFIA) is an airborne astronomical observatory comprised of a 2.5 meter infrared telescope mounted in the aft section of a Boeing 747SP aircraft that flies at operational altitudes between 37,000 and 45,00 feet, above 99\% of atmospheric water vapor. During routine operations, a host of instruments will be available to the astronomical community including cameras and spectrographs in the near- to far-IR; a sub-mm heterodyne receiver; and an high-speed occultation imager. One of the challenges for SOFIA (and all observatories in general) is providing a uniform set of tools that enable the non-expert General Investigator (GI) to propose, plan, and obtain observations using a variety of very different instruments in an easy and seamless manner. The SOFIA Data Cycle System (DCS) is an integrated set of services and user tools for the SOFIA Science and Mission Operations GI Program designed to address this challenge. Program activities supported by the DCS include:  
proposal preparation and submission by the GI; 
proposal evaluation by the telescope allocation committee and observatory staff;
 Astronomical Observation Request (AOR) preparation and submission by the GI;
 observation and mission planning by observatory staff;
 data processing and archiving;
 data product distribution.
In this poster paper we present an overview of the DCS concepts, architecture, and user tools that are (or soon will be) available in routine SOFIA operations. In addition, we present experience from the SOFIA Basic Science program, and planned upgrades.
\end{abstract}

\paragraph{Introduction}

The Data Cycle System (DCS) is designed primarily to support Science and Mission Operations activities associated with the General Investigator (GI) program for SOFIA. In general, observatory data flow is cyclic in nature (see Fig.~\ref{fig:DCS}).  The GI first develops and submits a proposal to the SOFIA Science Center (SSC) as part of the Phase~I planning process. Once his or her proposal is approved, the GI creates and updates Astronomical Observation Requests (AORs) during the Phase~II planning process. These AORs are then used in both flight planning and to generate scripts for on-aircraft execution with a facility Science Instrument (SI). After the AORs are completed in-flight, the resulting data are archived at the SSC and processed to produce higher-level data products which are then served back to the original GI via a simple user-interface---thus completing the cycle begun by the original proposal and perhaps driving new proposals. In addition, these data products are also made available to the astronomical community for archival research (after an initial proprietary period).  The goal of the DCS is to maximize the scientific productivity and efficiency of the observatory by providing a suite of easy-to-use tools and infrastructure that are integrated with each other as well as other applications at each step in the data cycle, and to provide a central ``hub'' for science and mission data at the SSC.

\articlefigure[scale=0.4]{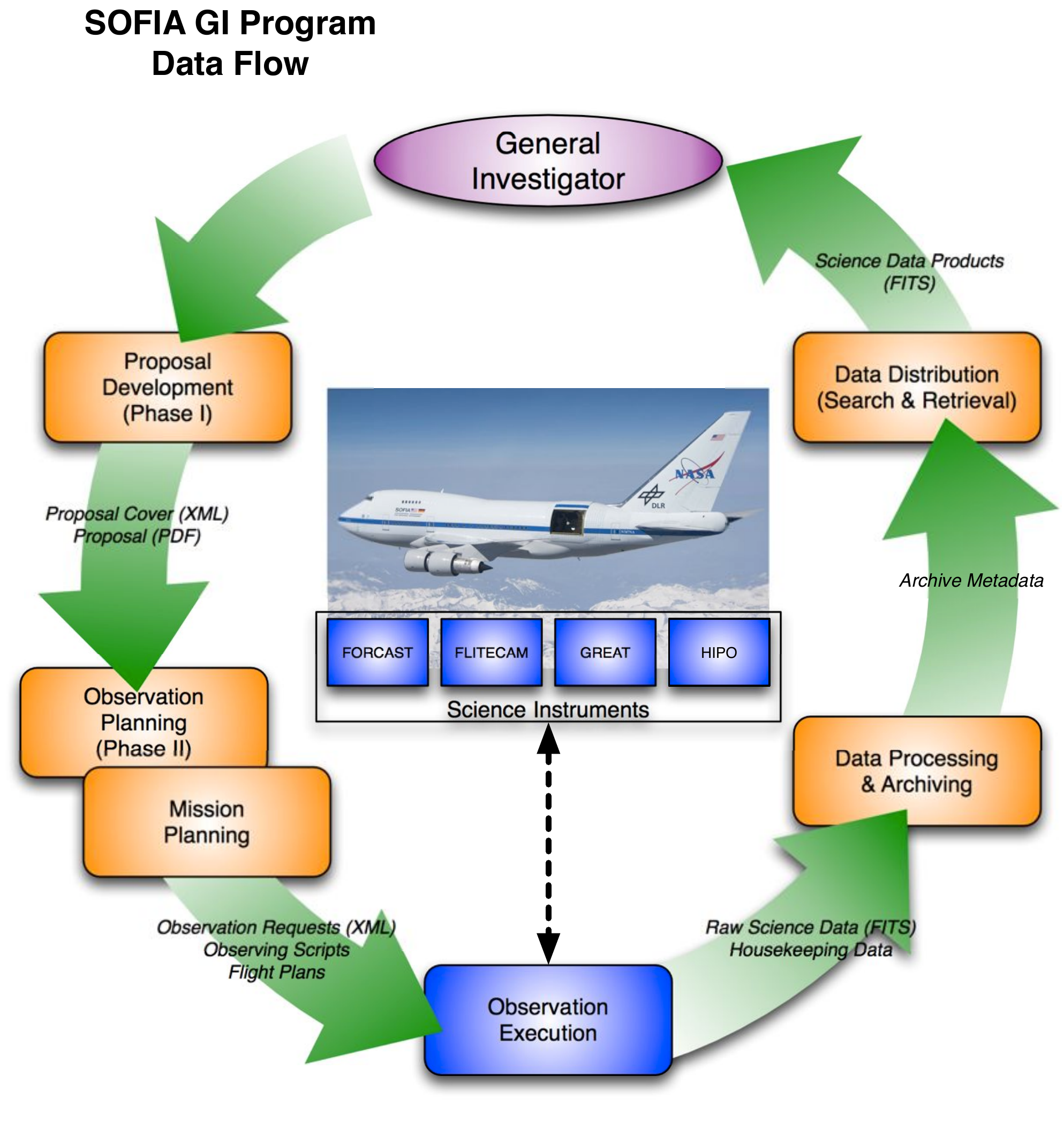}{fig:DCS}{SOFIA GI Program Data Cycle, showing the top-level processes and data flow to support GIs using the observatory.}

\paragraph{DCS Tools and Services}

User tools provided by the DCS\footnote{All DCS tools and functionality are accessed via the DCS web pages:  \url{http://dcs.sofia.usra.edu}. } 
include:

\begin{description}

\item[SOFIA Proposal Tool (SPT)] SOFIA proposals are created, modified, and submitted to the SSC using SPT, a stand-alone Java application available for a wide variety of platforms based on the Astronomers Proposal Tool (APT)  developed at STScI for HST.

\item[SOFIA-Spot] A stand-alone Java application that allows the GI to create, modify, and submit AORs for his or her observing program as part of the Phase~II planning process. S-SPOT provides a unified interface for configuring observations with the various SOFIA SIs; graphical overlays for visualizing an observation on a background image; and integrated access to existing IR survey mission archives (2MASS, MSX, IRAS, etc...).  S-Spot is based on the Spot tool developed for the Spitzer Space Telescope.

\item[SOFIA Instrument Time Estimator (SITE)] In order to assist in the creation of SOFIA observing proposals, the DCS web site provides an online time estimator that calculates the total integration time required or signal-to-noise achieved given specified atmospheric and instrument parameters. 

\item[Atmospheric transmission modeling] The DCS provides access to a web-based version of ATRAN~\citep{Lord:1992} that enables proposers to estimate the effects of atmospheric transmission for observations at varying altitudes, telescope elevations,  and water vapor overburdens.  

\item[Visibility Tool (VT)] Java applet that estimates target elevation and aircraft heading based on date, time, and aircraft location for a given astronomical target.  

\item[Archive Search and Retrieval] Web pages that provide search and retrieval capabilities on the DCS archive for both GIs and general users.  Files to be retrieved are bundled and staged on a webserver, where they can then be downloaded via sftp via a link sent to the user.

\end{description}

The DCS user tools are supported by a set of application, database, and filesystem servers at the SSC using a simple client/server architecture with all client traffic routed through webservers in a controlled DMZ (see Fig.~\ref{fig:DCS_Arch}).  Primary systems include:

\begin{description}

\item[Observation Planning System] The planning system is integrated with both SPT and SSpot to support the Phase~I and Phase~II planning activities.  Planning system data is organized around {\bf observing plans} (proposals), which are composed of {\bf AOR}s.  Planning data is made available to external clients ~\citep[e.g. Flight Management Infrastructure,][]{Gross:2009} using a well-defined API.  

\item[Archive System] The DCS Archive is the permanent repository for all raw scientific and housekeeping data accumulated during SOFIA flights and any processed data produced thereafter. Datasets from a flight are transferred to the SSC from the DAOF (where SOFIA is based) and then ingested into the DCS Archive where they are secured and parsed (as needed) to populate a summary database.  The archive enforces a 1-year proprietary period during which only associated GIs can access the science data for a particular observing plan.

\end{description}

\articlefigure[scale=0.33]{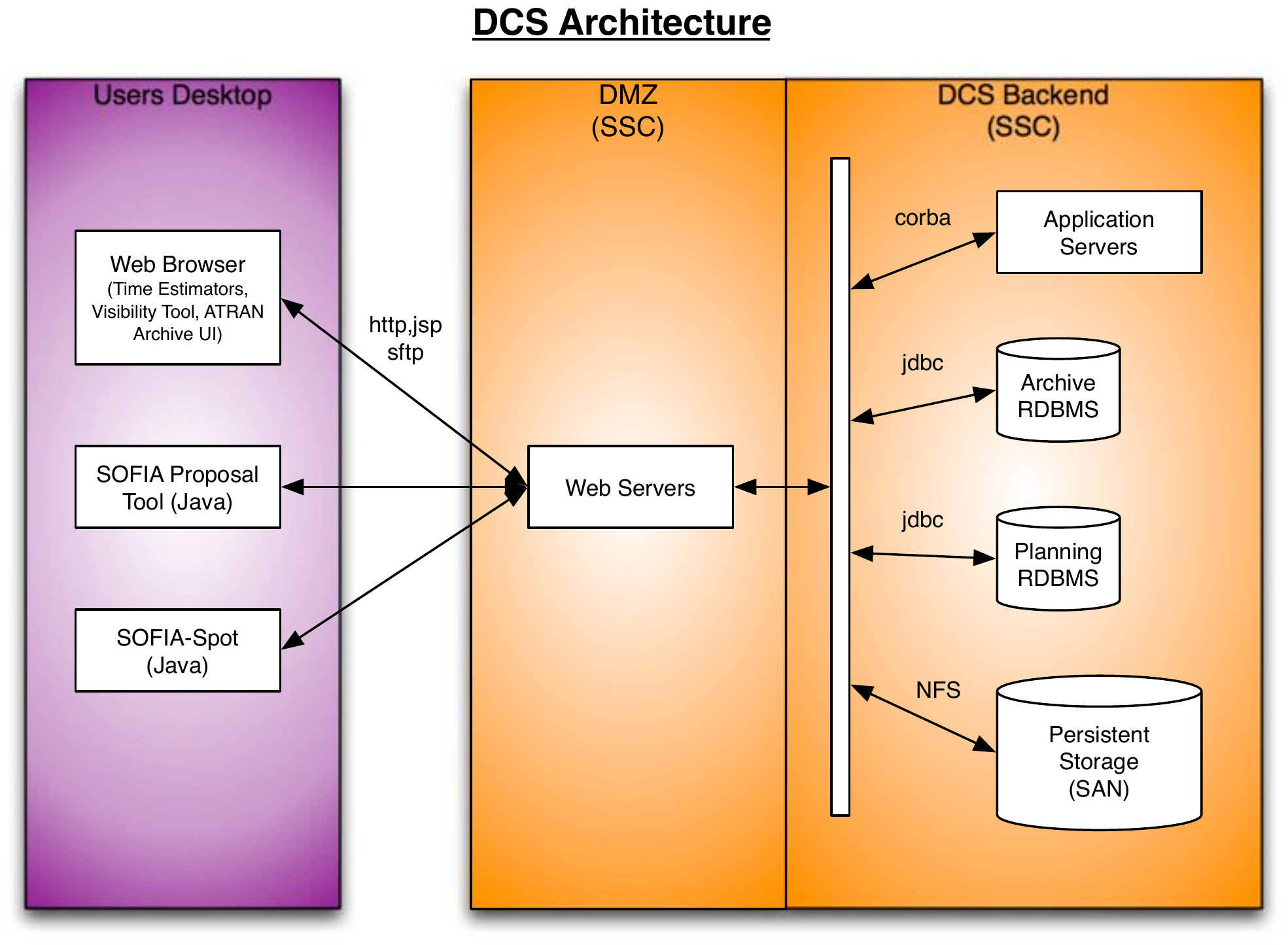}{fig:DCS_Arch}{High-level architectural diagram for the SOFIA Data Cycle System showing major system components.}

\paragraph{Utilization}

During 2011, 15 GI science flights were conducted to demonstrate the initial capabilities of SOFIA as part of the Basic Science Program~\citep{Young:2012}.  The DCS was used successfully for proposal submission and data archiving (both raw and reduced data products).  This was the first real-world use of the DCS and served as an excellent shake-down of the system:  a number of issues were reported by both observatory staff and outside GIs that have been fixed in preparation for routine operations.

Late in 2011, the SSC issued the first general Call for Proposals for Cycle~1---a full year of flights with $> 200$~hours of observing time available to the US and German astronomical communities.\footnote{
\url{http://www.sofia.usra.edu/Science/proposals/cycle1/index.html}}
Over 170 proposals were submitted during Phase~I using SPT.  S-Spot was successfully used during Phase~II planning by the Cycle~1 GIs for creating and submitting AORs (a process that was done interactively with the GIs during Basic Science).  There are now $\approx 800$ AORs  stored in the DCS that are currently being used by the flight scheduling and planning tools.

The DCS Archive currently contains raw and reduced data products for $> 300$ observations ($\approx 232$~GB) carried out during Basic Science. In addition, the DCS archive contains over 850~GB of mission data (guide camera images, housekeeping data, etc...) and other ancillary data products (mission audio, logs, etc...).  

\paragraph{Planning for the Future}

Current loads on the DCS are light.  But the number of flight hours available will increase substantially in the coming years, which in turn means higher user loads and demands on the system.  The DCS backend application servers employ a distributed architecture that can be scaled-up as needed, and increased web traffic can be handled by adding additional webservers behind a load-balancer.  

The DCS must also support changes from observing cycle to observing cycle resulting from instrument upgrades.  The current set of user tools stores parameters associated with each instrument in easy-to-update configuration and template files:  As instrument filters, sensitivities, and other parameters change, the tools can be rapidly modified, verified, and released to the SOFIA user community.

\acknowledgements 
\small
The SOFIA Data Cycle System was initially developed under cooperative agreement between NASA and the Universities Space Research Association (USRA), in conjunction with the Rochester Institute of Technology and the University of California, Los Angeles. Development has continued under NASA contract NAS2-97001 to USRA. The authors would like to acknowledge  all the dedicated scientists and engineers who have contributed to the development of the DCS over the years, including:  Francis J.~Nelbach, 
Goeran Sandell, 
Neill Callis, 
Jonathan Adams, 
Miguel Charcos-Llorens, 
Rosemary Alles,
Kaori Nishikida,
Erich Proudfit, 
Robert Perez,
Elizabeth Moore,
Prof.~Harvey Rhody, 
Patrick Stein,
Scott Lawrence, 
William Hoagland,
Prof.~Mark Morris, 
J.~Milburn,
Sean Colgan,
and David Goorvitch.

\bibliographystyle{asp2010.bst}

\bibliography{Shuping_P50_REV1}

\end{document}